\documentclass{ws-procs9x6}

\begin{document}

\title{The high-energy limit of inclusive and diffractive deep inelastic 
scattering in QCD}

\author{Cyrille Marquet}

\address{Service de Physique Th\'eorique, CEA/Saclay\\
91191 Gif-sur-Yvette cedex, France\\
E-mail: marquet@spht.saclay.cea.fr}

\maketitle

\abstracts{
Following recent progresses in the understanding of high-energy scattering in 
QCD, we derive the first phenomenological consequences of Pomeron loops, in the 
context of both inclusive and diffractive deep inelastic scattering. In 
particular, we discuss diffusive scaling, a new scaling law that emerges for 
sufficiently high energies and up to very large values of $Q^2,$ well above the 
proton saturation momentum.}

\section{Introduction}

The Good-and-Walker picture of diffraction was originally meant to describe soft 
diffraction. They express an hadronic projectile $|P\rangle\!=\!\sum_n 
c_n|e_n\rangle$ in terms of hypothetic eigenstates of the interaction with the 
target $|e_n\rangle,$ that can only scatter elastically:
$\hat{S}|e_n\rangle\!=\!(1\!-\!T_n)|e_n\rangle.$ The total, elastic and 
diffractive cross-sections are then easily obtained:
 \begin{equation}
\sigma_{tot}=2\sum_n c_n^2 T_n\hspace{0.5cm}
\sigma_{el}=\Big[\sum_n c_n^2 T_n\Big]^2\hspace{0.5cm}
\sigma_{diff}=\sum_n c_n^2 T_n^2\ .
\end{equation}
 
It turns out that in the high energy limit, there exists a basis of eigenstates 
of the large$-N_c$ QCD $S-$matrix: sets of quark-antiquark dipoles 
$|e_n\rangle\!=\!|d(r_1),\dots,d(r_n)\rangle$ caracterized by their transverse 
sizes $r_i.$ In the context of deep inelastic scattering (DIS), we also know the 
coefficients $c_n$ to express the virtual photon in the dipole basis. For 
instance, the equivalent of $c_1^2$ for the one-dipole state is the photon 
wavefunction $\phi(r,Q^2).$ 

This realization of the Good-and-Walker picture allows to write down an exact 
(within the high-energy and large$-N_c$ limits) factorization 
formula~\cite{difscal} for the diffractive cross-section in DIS in terms of 
dipole scattering amplitudes off the target proton, such as $\left\langle 
T(r_1)\dots T(r_n)\right\rangle_Y.$ The average $\langle\ .\ \rangle_Y$ is an 
average over the proton wavefunction which gives the energy dependence to the 
cross-section ($Y\!=\!\log(1/x)\!\sim\!\log(s)$ is the rapidity). 

\section{The geometric and diffusive scaling regimes}

Within the high-energy and large$-N_c$ limits, the dipole amplitudes are 
obtained from the Pomeron-loop equation~\cite{ploop} derived in the leading 
logarithmic approximation in QCD. This is a Langevin equation which exhibits the 
stochastic nature~\cite{fluc} of high-energy scattering processes in QCD. Its 
solution is an event-by-event dipole scattering amplitude function of 
$\rho\!=\!-\log(r^2Q_0^2)$ and $Y$ ($Q_0$ is a scale provided by the initial 
condition). It is characterized by a saturation scale $Q_s$ which is a random 
variable whose logarithm is distributed according to a Gaussian probability 
law~\cite{prob}. The average value is $\log(\bar{Q}_s^2/Q_0^2)\!=\!\lambda Y$ 
and the variance is $\sigma^2\!=\!DY$ (see Fig.1). The dispersion coefficient 
$D$ allows to distinguish between two energy regimes: the geometric scaling 
regime ($DY\!\ll\!1$) and diffusive scaling regime ($DY\!\gg\!1$).

The following results for the averaged amplitude will be needed to derive the 
implications for inclusive and diffractive DIS:
\begin{eqnarray}
\left\langle T(r_1)\dots T(r_n)\right\rangle_Y&\stackrel{Y\ll1/D}{=}&
\left\langle T(r_1)\right\rangle_Y\dots\left\langle T(r_n)\right\rangle_Y\ ,\\
\left\langle T(r_1)\dots T(r_n)\right\rangle_Y&\stackrel{Y\gg1/D}{=}&
\left\langle T(r_<)\right\rangle_Y,\hspace{0.3cm}r_<=\min(r_1,\dots,r_n)\ .
\end{eqnarray}
All the scattering amplitudes are expressed in terms of $\langle T(r)\rangle_Y,$ 
the amplitude for a single dipole which features the following scaling 
behaviors:
\begin{eqnarray}
\left\langle 
T(r)\right\rangle_Y\stackrel{Y\ll1/D}{\equiv}T_{gs}(r,Y)&=&\displaystyle
T\left(r^2\bar{Q}_s^2(Y)\right)\ ,\\
\left\langle 
T(r)\right\rangle_Y\stackrel{Y\gg1/D}{\equiv}T_{ds}(r,Y)&=&\displaystyle
T\left(\frac{\log(r^2\bar{Q}_s^2(Y))}{\sqrt{DY}}\right)\ .
\end{eqnarray}

\begin{figure}[b]
\begin{minipage}[t]{41mm}
\centerline{\epsfxsize=4.1cm\epsfbox{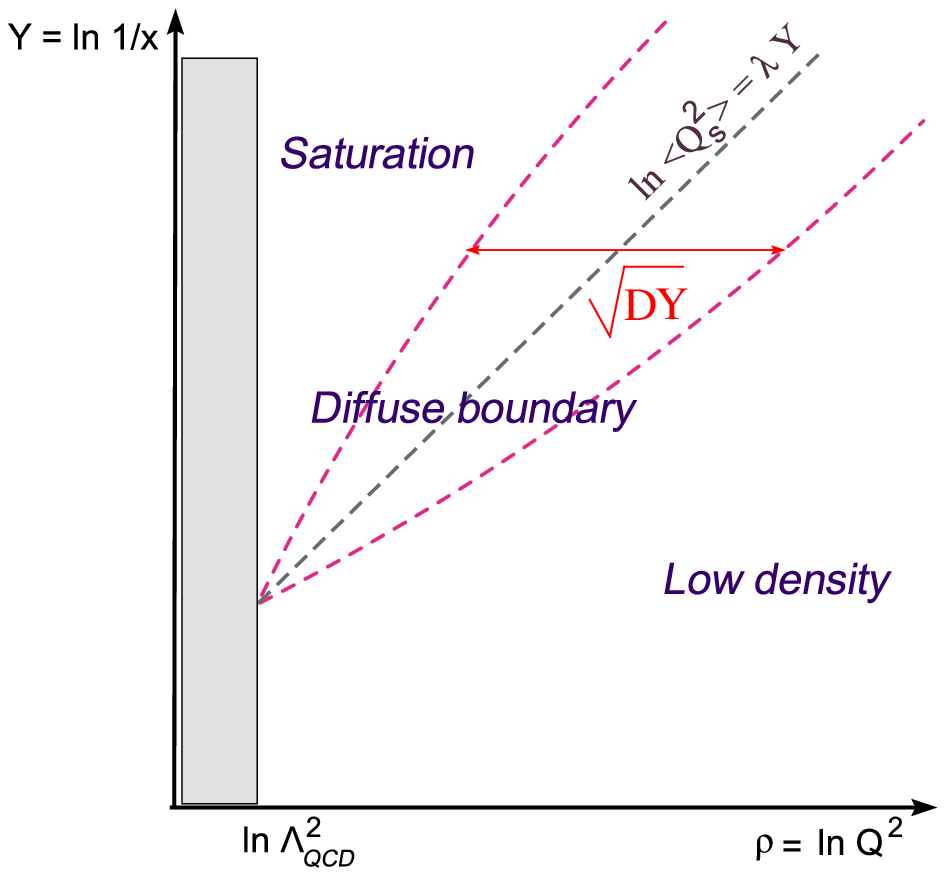}}
\end{minipage}
\hspace{\fill}
\begin{minipage}[t]{71mm}
\vspace{-3.9cm}\centerline{\epsfxsize=3.9cm\rotatebox{-90}{\epsfbox{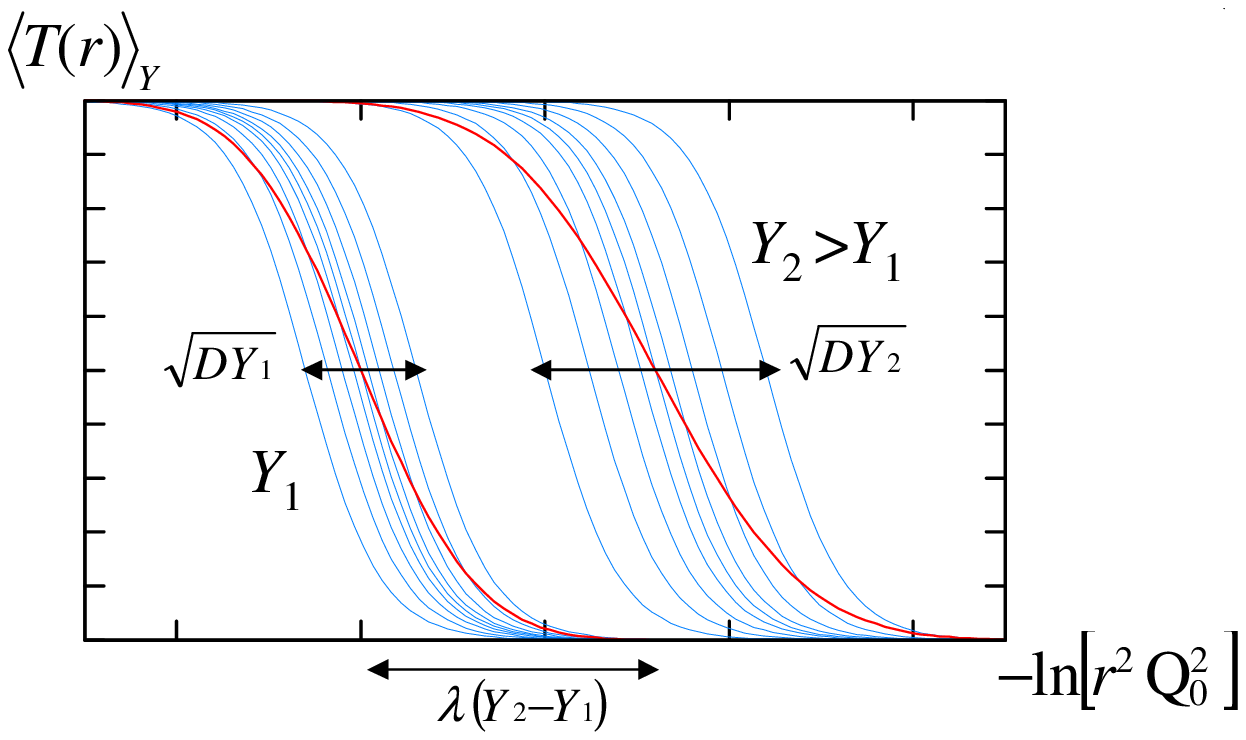}}}
\end{minipage}
\caption{Left plot: a diagram representing the stochastic saturation line in the 
$(\rho,Y)$ plane, the diffusive saturation boundary is generated by the 
evolution. Right plot: different realizations of the event-by-event scattering 
amplitude (gray curves) and the resulting averaged physical amplitude $\langle 
T(r)\rangle$ (black curve) as a function of $\rho$ for two different values of 
$Y$ in the diffusive scaling regime.}
\end{figure}

\section{Implications for inclusive and diffractive DIS}

We shall concentrate on the diffusive scaling regime, in which the dipole 
scattering amplitude can be written as follows~\cite{errorf} for 
$-\log(r^2\bar{Q}_s^2(Y))\!\ll\!DY:$
\begin{equation}
T_{ds}(r,Y)=\frac12 
Erfc\left(-\frac{\log(r^2\bar{Q}_s^2(Y))}{\sqrt{DY}}\right)\ .
\label{erfc}\end{equation}
From this, one obtains the following analytic estimates \cite{difscal} for the 
$\gamma^*\!-\!p$ total cross-section in DIS and for the diffractive one 
integrated over $\beta$ from $\beta_<:$
\begin{eqnarray}
\frac{d\sigma_{tot}}{d^2b}(x,Q^2)&=&\frac{N_c\alpha_{em}}{12\pi^2}\sum_f e_f^2
\sqrt{\pi D\log(1/x)}\ \frac{e^{-Z^2}}{Z^2}\ ,\\
\frac{d\sigma_{diff}}{d^2b}(x,Q^2,\beta_<)&=&\frac{N_c\alpha_{em}}{48\pi^2}
\sum_f e_f^2 \sqrt{D\log(1/x)}\ \frac{e^{-2Z^2}}{Z^3}\ .
\end{eqnarray}
The variable $Z$ is reminiscent of the scaling variable of the dipole amplitude: 
\begin{equation}
Z=\frac{\log(Q^2/\bar{Q}_s^2(x))}{\sqrt{D\log(1/x)}}
\end{equation}
and shows that in the diffusive scaling regime, both inclusive and diffractive 
scattering are dominated by small dipole sizes $r\!\sim\!1/Q.$ Also the 
cross-sections do not feature any Pomeron-like (power-law type) increase with 
the energy and the diffractive cross-section (which does not depend on 
$\beta_<$) is dominated by the scattering of the quark-antiquark ($q\bar q$) 
component, corresponding to $\beta\lesssim1.$ These features a priori expected 
in the saturation regime ($Q^2<\bar{Q}_s^2$) are valid up to values of $Q^2$ 
much bigger than $\bar{Q}_s^2:$ in the whole diffusive scaling regime for 
$\log(Q^2/\bar{Q}_s^2(Y))\!\ll\!DY$ (see Fig.2).

\begin{figure}[h!]
\centerline{\epsfxsize=3.8cm\epsfbox{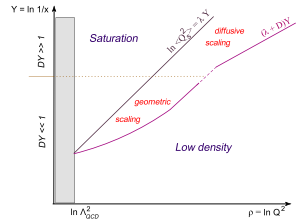}}
\caption{A phase diagram for the high-energy limit of inclusive and diffractive 
DIS in QCD. Shown are the average saturation line and the approximate boundaries 
of the scaling regions at large values of $\rho\!\sim\!\ln Q^2.$ With increasing 
$Y,$ there is a gradual transition from geometric scaling at intermediate 
energies to diffusive scaling at very high energies.}
\end{figure}

\clearpage

\begin{figure}[t]
\centerline{\epsfxsize=4cm\rotatebox{-90}{\epsfbox{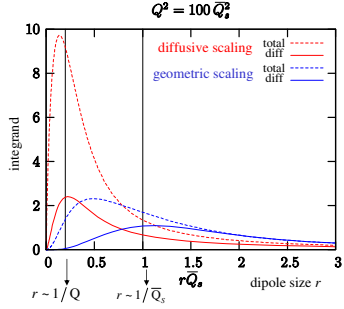}}}
\caption{The integrands of \eqref{dis} plotted as a function of $r\bar{Q}_s$ and 
computed with
two expressions for the dipole amplitude: in the geometric and 
diffusive scaling regimes.}
\end{figure}

The inclusive cross-section and the $q\bar q$ contribution to the diffractive 
one are obtained from the dipole amplitude $\langle T(r)\rangle_Y$ in the 
following way:
\begin{equation}
\frac{d\sigma_{tot}}{2\pi\ d^2b}=\int dr^2 \Phi(r,Q^2)\langle T(r)\rangle_Y\ ,\, 
\frac{d\sigma_{diff}}{\pi\ d^2b}=\int dr^2 \Phi(r,Q^2)\langle T(r)\rangle^2_Y\ .
\label{dis}\end{equation}

In order to better exhibit the dominance of small dipole sizes $r\!\sim\!1/Q$, 
we represent in Fig.3 the integrands of \eqref{dis} as a function of the dipole 
size $r.$ Keeping $Q/\bar{Q}_s\!=\!10$ fixed, we use \eqref{erfc} in the 
diffusive scaling regime and $T_{gs}(r,Y)\!=\!1\!-\!e^{-r^2\bar{Q}_s^2(Y)/4}$ in 
the geometric scaling regime. The difference between the geometric and diffusive 
scaling is striking. For the latter, both inclusive and diffractive scattering 
are dominated by inverse dipole sizes of the order of the hardest infrared 
cutoff in the problem: the hardest fluctuation of the saturation scale, which is 
as large as $Q.$

In the diffusive scaling regime, up to values of $Q^2$ much bigger than the 
saturation scale $\bar{Q}_s^2$, cross-sections are dominated by rare events in 
which the photon hits a black spot that he sees at saturation at the scale 
$Q^2.$ In average the scattering is weak, but saturation is the only relevant 
physics.

\end{document}